\documentclass[11pt]{article}
\usepackage{moriond,epsfig}

\def\be{\begin{equation}}
\def\ee{\end{equation}}
\def\bea{\begin{eqnarray}}
\def\eea{\end{eqnarray}}

\begin{document}
\vspace*{4cm}
\title{Measurement of the W mass at LEP and recent Results of the Electroweak Fit}

\author{ CHRISTIAN ROSENBLECK }

\address{ III. Physikalisches Institut A, RWTH Aachen, D-52062 Aachen, Germany }

\maketitle\abstracts{
  The measurement of the mass of the W boson, $m_{\rm W}$, by direct reconstruction is described.
  The combination of all measurements of the four LEP experiments yields $m_{\rm W} = 80.412 \pm 0.042$ GeV.
  Together with several other measurements the extracted $m_{\rm W}$ value enters the electroweak fit,
  which tests the Standard Model (SM) predictions and extracts SM parameters which have not yet been measured.
  The latest results of the fit are presented.
}

\section{Introduction}
The four LEP experiments ALEPH, DELPHI, L3, and OPAL have selected about 40,000 W boson pairs at
centre-of-mass energies above the production threshold, corresponding to a luminosity of 0.7 fb$^{-1}$ per experiment.
The W bosons decay into quarks or leptons, which are reconstructed to measure the mass of the W boson, $m_{\rm W}$.

\section{W boson mass extraction}
\begin{figure}
  \begin{center}
    \includegraphics[width=0.45\textwidth]{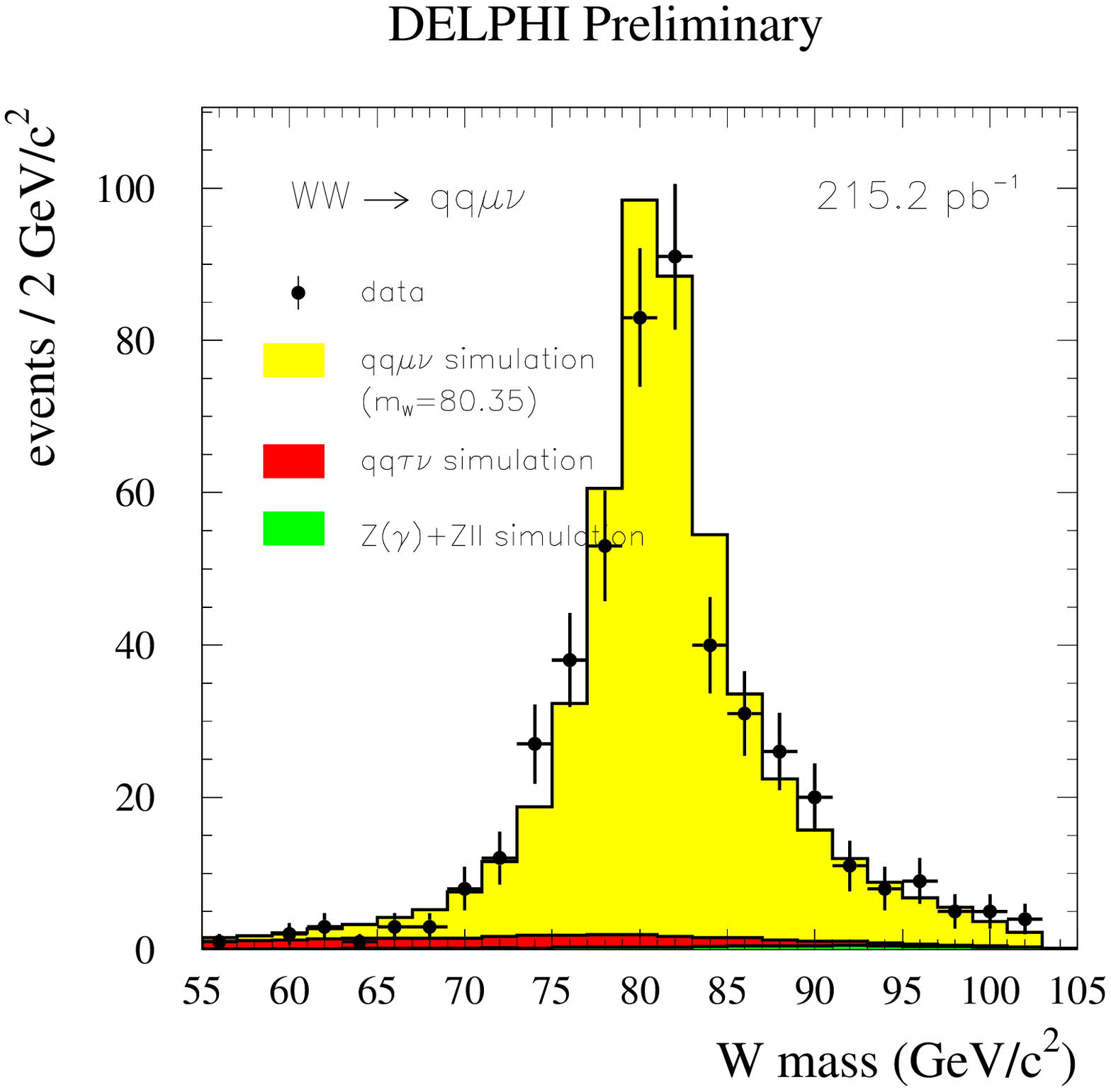}
    \hfill
    \includegraphics[width=0.45\textwidth]{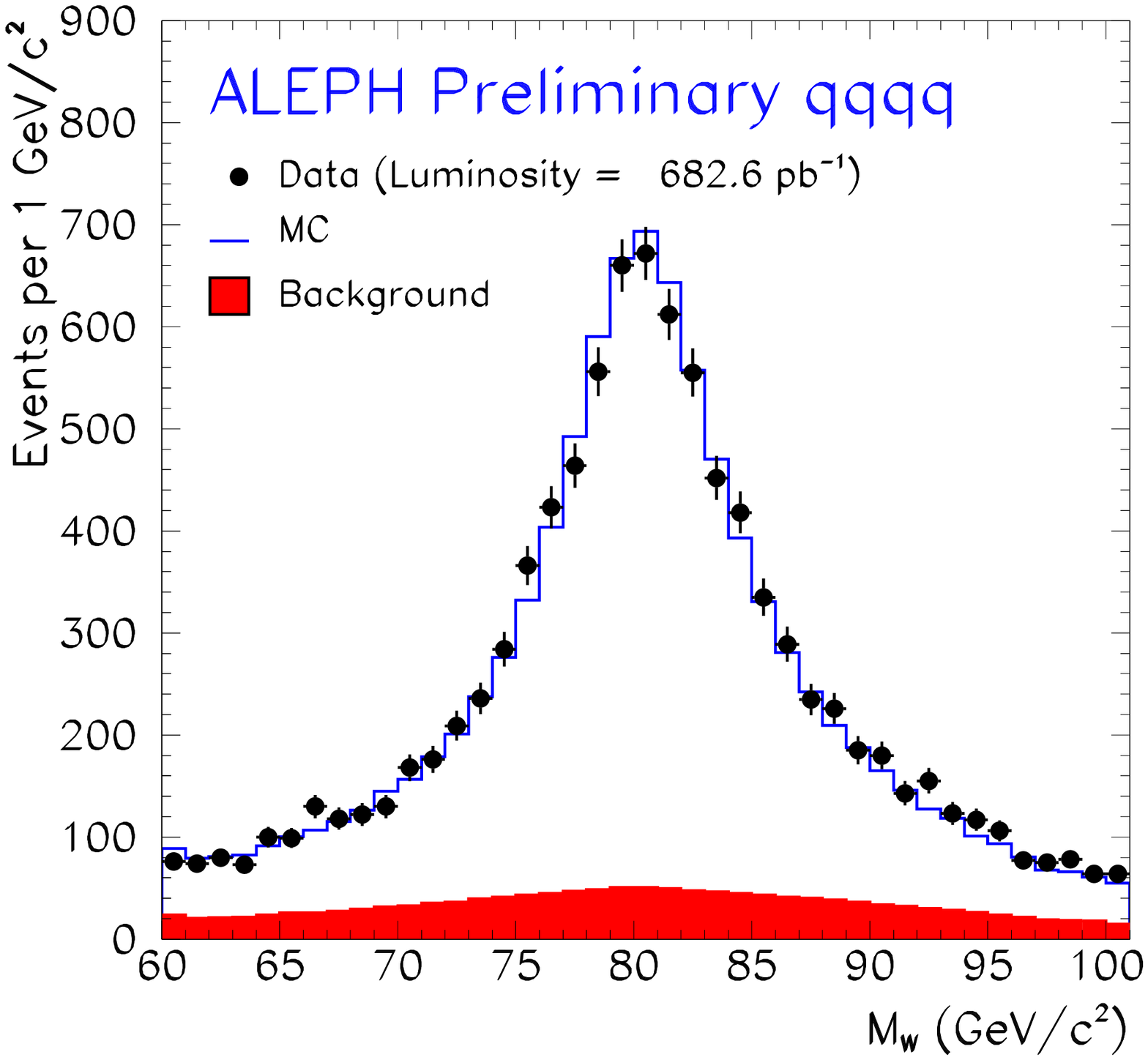}
  \end{center}
  \vspace{-22pt}
  a) \hspace{.49\textwidth}b)
  \caption{
    Spectra of the effective mass, reconstructed from a) $\rm W^{+}W^{-}\rightarrow qq\mu\nu$
    and b) $\rm W^{+}W^{-}\rightarrow qqqq$ events.
  }
  \label{fig:mwspectra1}
\end{figure}

\begin{figure}
  \begin{center}
    \includegraphics*[width=0.9\textwidth,clip=]{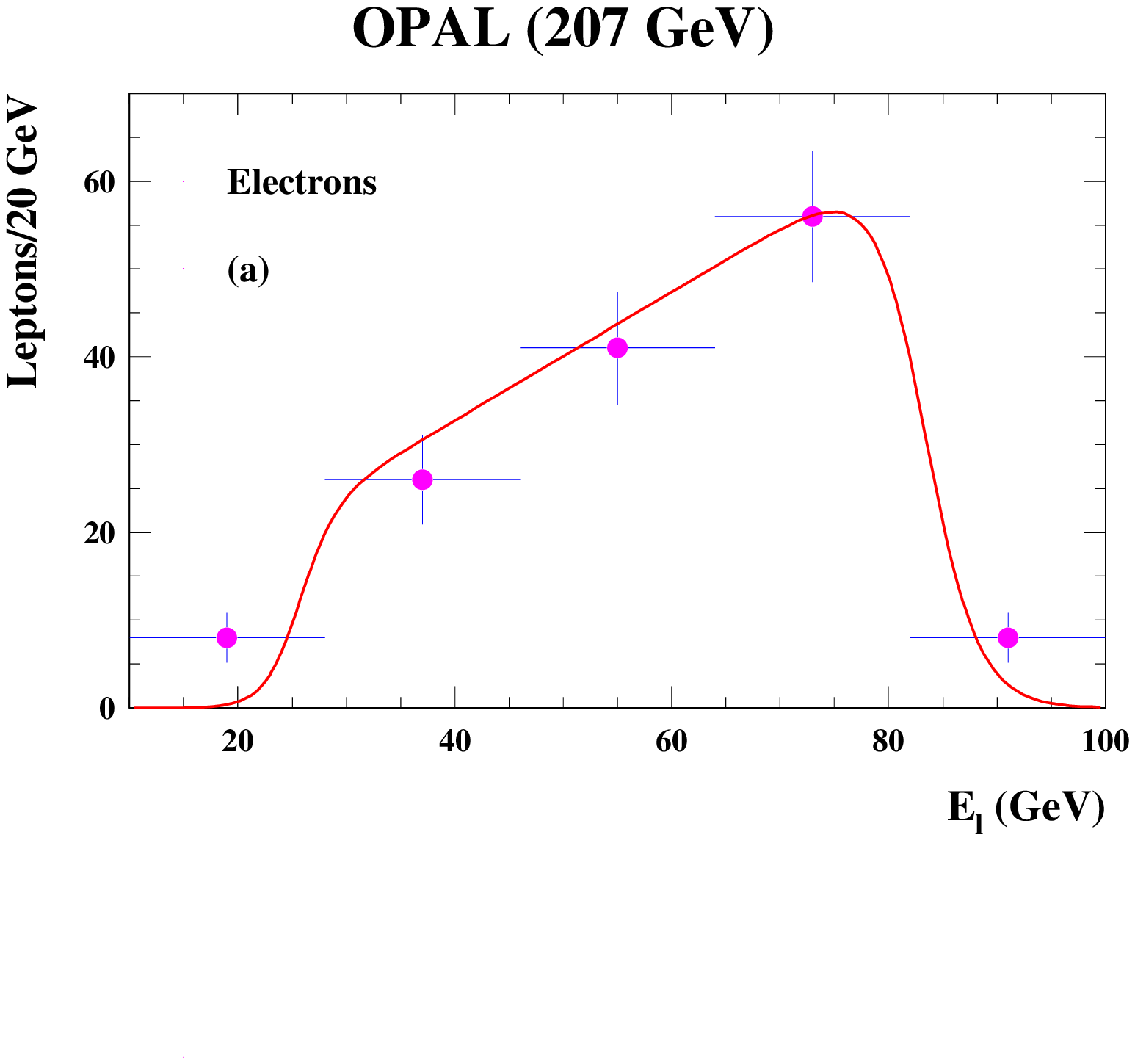}
  \end{center}
  \caption{
    Distribution of the lepton energy in $\rm W^{+}W^{-}\rightarrow e^{+}\nu e^{-}\nu$ events.
  }
  \label{fig:elep}
\end{figure}

At energies above the kinematic threshold, the W boson mass is extracted from the spectrum of the effective mass
reconstructed in $\rm W^{+}W^{-}\rightarrow qqqq$ and $\rm W^{+}W^{-}\rightarrow qq\ell\nu$ events~\cite{mw},
where the effective mass is reconstructed from the W decay particles.
To improve the mass resolution, a kinematic fit is applied to the measured angles and energies,
conserving the four-momentum of the events.
Additionally, an equal-mass constraint is applied. 
In qqqq(g) events the assignment of the jets to the W bosons is ambiguous.
In case of four jets there exist three different pairings, if an additional gluon jet is allowed, 
there are ten possible assignment.
The W-boson mass is extracted from the spectra of the effective mass, as shown in Figure~\ref{fig:mwspectra1} 
for $\rm qq\mu\nu$ and qqqq events.
These spectra are based on Breit-Wigner functions.
They are however distorted by hadronisation effects, initial state radiation and detector effects.
Different methods to extract the W boson mass from these spectra are applied:
A Breit-Wigner function is fitted to the data spectrum correcting a possible bias by Monte Carlo studies,
a likelihood is constructed by convoluting the Breit-Wigner function with ISR and detector resolution,
or the Monte Carlo events are re-weighted to obtain Monte Carlo samples representing different W-boson masses
which are fitted to the data spectrum.
In the $\ell\nu\ell\nu$ channel there are at least two neutrinos that remain undetected and carry away information.
Therefore, the energy of the charged leptons and a pseudo-mass variable are used to determine 
$m_{\rm W}$~\cite{Abbiendi:2002ay}.
As an example, the lepton energy spectrum in $\rm e^{+}\nu e^{-}\nu$ events is shown in Figure~\ref{fig:elep}.

\section{Systematic Uncertainties}
Table~\ref{tab:syst} summarizes the systematic uncertainties on the measurement of the W-boson mass used in the 
latest LEP wide combination~\cite{lepwcombi}.
\begin{table}[b]
  \begin{center}
    \begin{tabular}{l|c|c|c}
                 & \multicolumn{3}{c}{Effect [MeV]} \\
      Source     & $\rm qq\ell\nu$ & qqqq & ffff    \\
      \hline
      LEP energy & 17              & \phantom{1}17   & 17      \\
      Detector effects   & 14      & \phantom{1}10   & 14      \\
      ISR/FSR    & \phantom{0}8    & \phantom{10}8 & \phantom{0}8 \\
      Hadronisation & 19           & \phantom{1}18   & 18      \\
      Bose-Einstein correlations & -- & \phantom{1}35 & \phantom{0}3\\
      Coulour reconnection & --    & \phantom{1}90   & \phantom{0}9 \\
      Other      & \phantom{0}4    & \phantom{10}5 & \phantom{0}4 \\
      \hline
      Total      & 31              & 101     & 31
    \end{tabular}
  \end{center}
  \caption{
    Systematic uncertainties of the W mass measurement.
  }
  \label{tab:syst}
\end{table}

The estimation of the LEP beam energy uncertainty used in the preliminary combination (yielding a W mass uncertainty of 
17 MeV) has recently been improved significantly~\cite{Assmann:2004gc}.
Therefore, the final W-mass uncertainty originating from LEP beam energy measurements will account to 10~MeV in a final
combination.

Detector systematics include uncertainties of the measurement of angles and energies.
Monte Carlo simulations of the detector response are calibrated using data taken at calibration runs on the Z pole.
The systematic uncertainty due to detector effects amounts to 14 MeV in the combination of all channels.

Effects of the simulation of photon radiation in the initial and final state are studied by re-weighting Monte Carlo
events to different orders of $\alpha$.
Another estimation of these effects can be achieved by a comparison of the two Monte Carlo generators 
{\sc Racoon}~\cite{Denner:2002cg} and {\sc YFSWW3}~\cite{Jadach:2001uu}.
As larger differences are seen compared to the re-weighting approach, this contribution to the systematic uncertainty 
might increase in future.

Since the hadronisation of quarks is a non-perturbative process, only phenomenological models exist.
From comparisons of different hadronisation models ({\sc Ariadne}~\cite{Lonnblad:1992tz}, {\sc Herwig}~\cite{herwig}, 
and {\sc Pythia}~\cite{Sjostrand:2000wi}), an uncertainty on $m_{\rm W}$ of 18 MeV is estimated.

Final state interactions (FSI), namely Bose-Einstein correlations (BEC) and Colour Reconnection (CR),
would lead to a cross-talk between the decay products of different W bosons.
Therefore, the existence of FSI in data could introduce large biases in the determination of the W-boson mass in the 
four-jet channel.
The uncertainties due to these effects drastically decrease the weight of the four jet channel in the combination 
of the $\rm qq\ell\nu$ and qqqq channels. 

\begin{figure}
  \begin{center}
    \begin{minipage}{0.45\textwidth}
      \includegraphics[width=\textwidth]{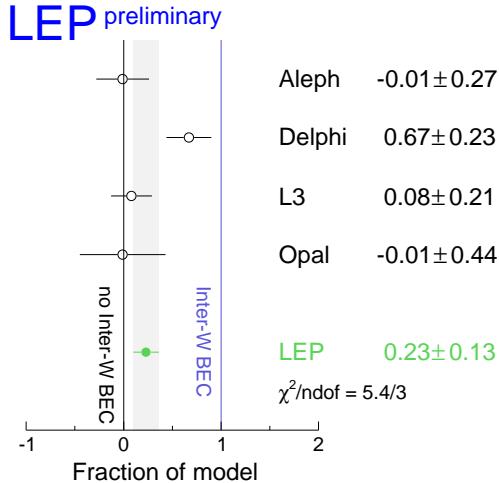}
    \end{minipage}
  \end{center}
  \caption{
    Combination of preliminary measurements of the fraction of the BEC model for each LEP experiment.
  }
  \label{fig:bec}
\end{figure}

BEC lead to the enhanced production of identical bosons close to each other in phase space.
They were studied in dedicated analyses~\cite{becanal}, comparing four-jet events with $\rm qq\ell\nu$ events.
While ALEPH, L3 and OPAL have published analyses establishing BEC between bosons originating from the same W,
BEC between bosons from different W bosons (``Inter-W BEC'') are disfavoured.
DELPHI favours moderate BEC between bosons from different Ws.
Figure~\ref{fig:bec} shows the fraction of the model seen for the latest combination~\cite{lepcombi}, 
which does not include the most recent results.
The preliminary combination allows at maximum 36\% of the effect of the BEC model.
Applying this fraction to the shift observed with the full model decreases the systematic uncertainty
quoted in Table~\ref{tab:syst} to about 13 MeV.

\begin{figure}
  \begin{center}
    \begin{minipage}{0.45\textwidth}
      \includegraphics[width=\textwidth]{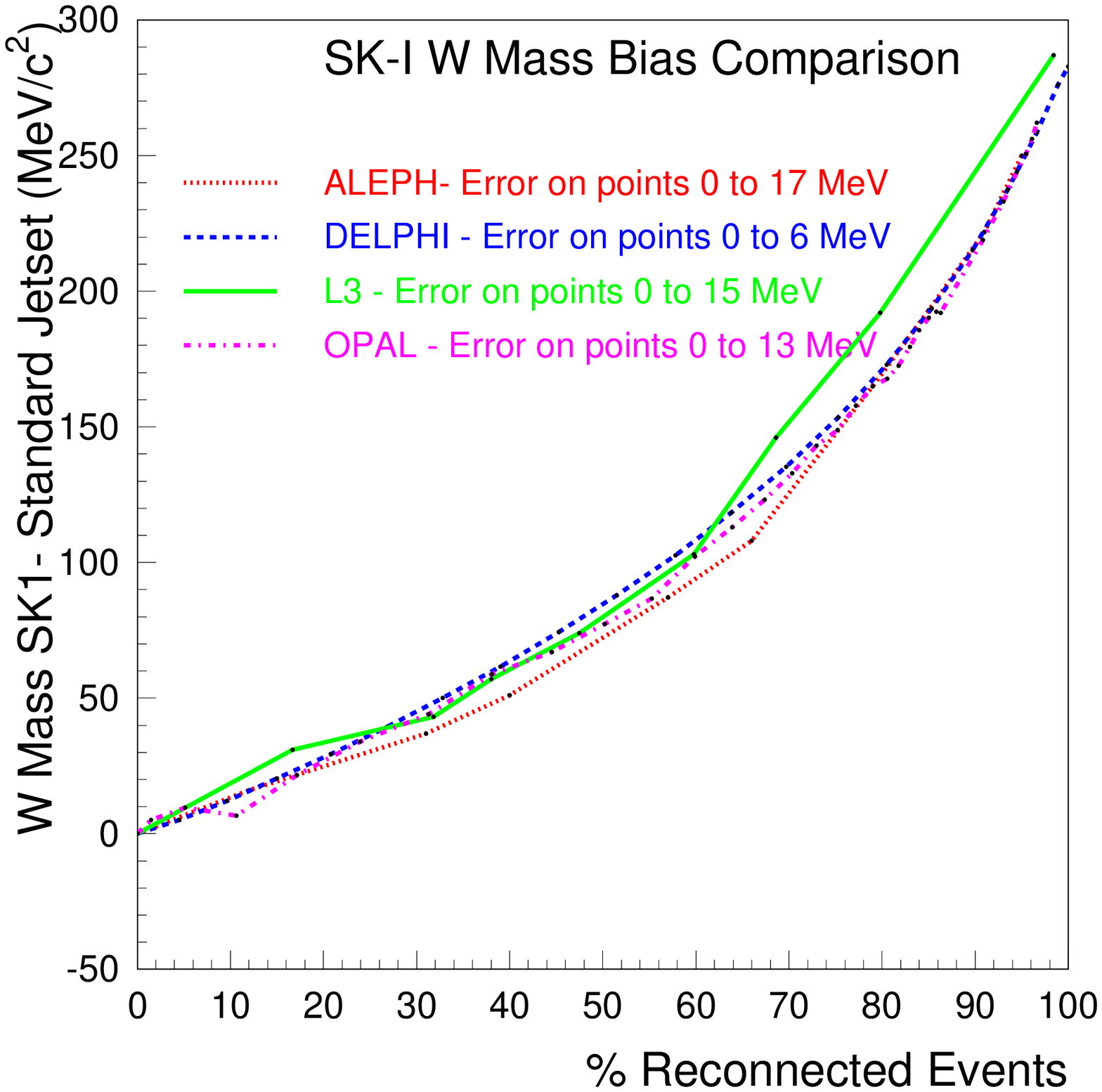}
    \end{minipage}
    \hfill
    \begin{minipage}{0.45\textwidth}
      \includegraphics[width=\textwidth]{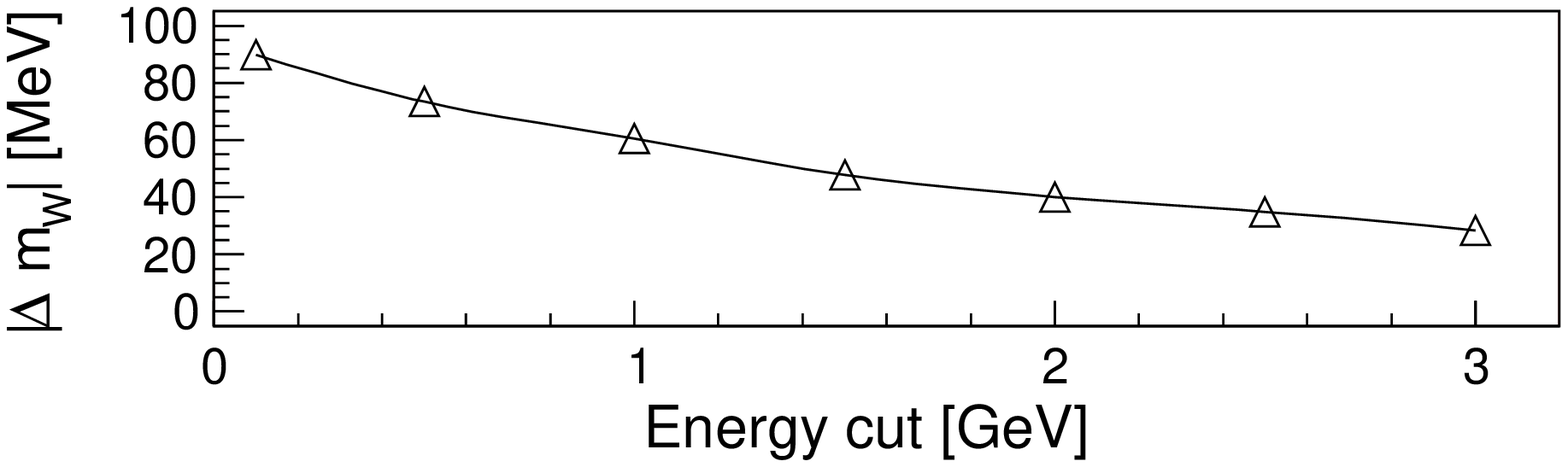}

      \includegraphics[width=\textwidth]{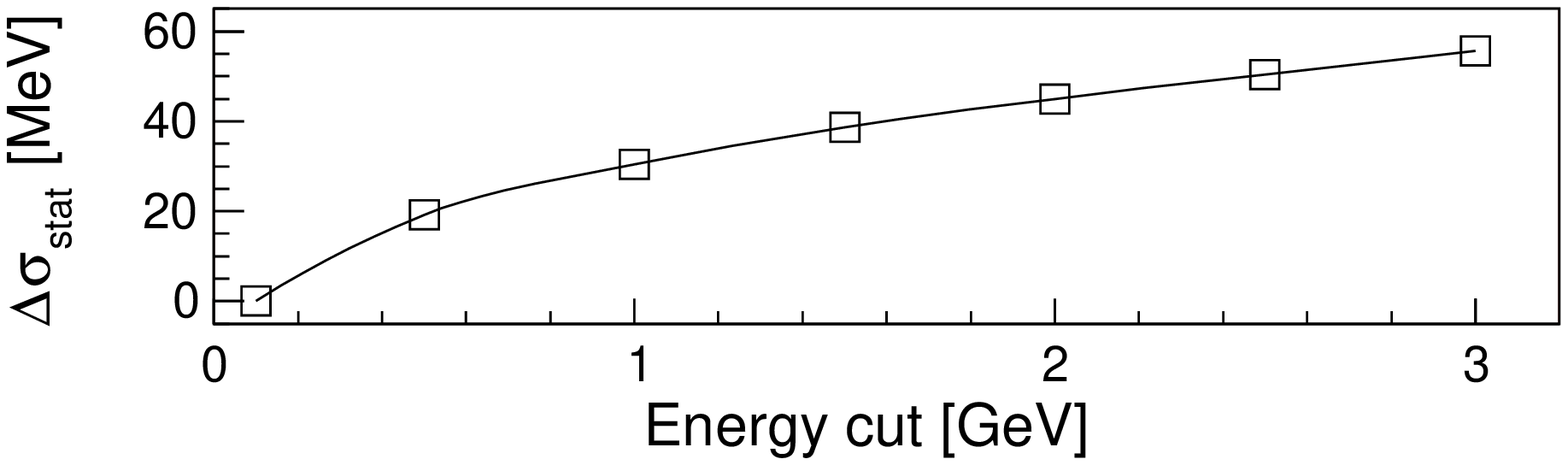}

      \includegraphics[width=\textwidth]{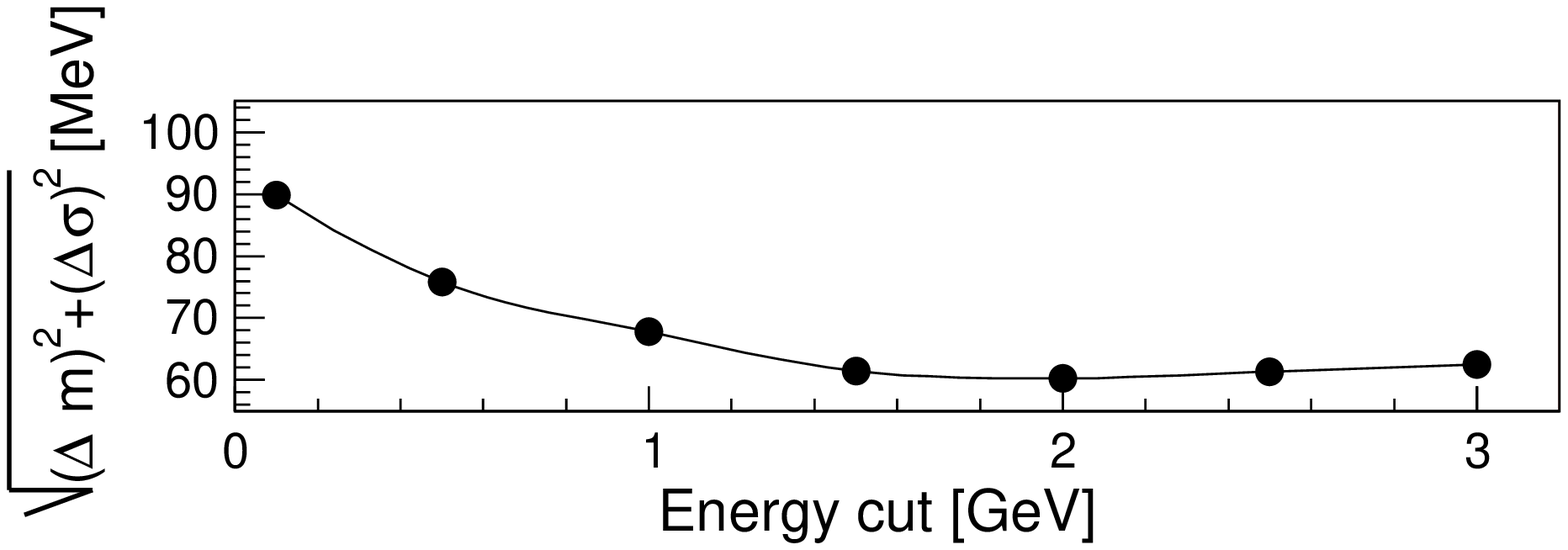}
    \end{minipage}
  \end{center}
  \vspace{-22pt}
  a) \hspace{.49\textwidth}b)
  \vspace{22pt}
  \caption{
    a) Shift of the extracted W mass when a certain fraction of Monte Carlo events contains Colour Reconnection effects.
    b) Effects of applying a cut on the cluster energy when clustering jets. See text for more details.
  }
  \label{fig:crshift}
\end{figure}

Colour Reconnection describes the possible exchange of momentum between particles from different W bosons 
due to reorganisation of the colour flow in the events.
This was first proposed by Gustafsson, Petersson and Zerwas~\cite{Gustafson:1988fs}.
CR is modelled in the Monte Carlo generators {\sc Ariadne}, {\sc Herwig}, and {\sc Pythia}.
The SKI model~\cite{Khoze:1998gi} of Pythia gives the largest effects.
This model contains a free parameter that tunes the strength of the reconnection.
Figure~\ref{fig:crshift} shows the shift of the extracted W-boson mass with respect to the fraction of reconnected events.
Dedicated analyses~\cite{crlep} try to measure this parameter by studying the particle flow between
jets from the same W and different W bosons, or by Cone/Cut studies as explained below.
The latest results have not yet been combined; the combination of preliminary measurements~\cite{crcombi} yields 
an upper limit of 2.13 for the free parameter.
This upper limit leads to an uncertainty of 90 MeV as quoted in Table~\ref{tab:syst}.

This possible bias could be reduced by using different jet clustering algorithms.
Since low energy clusters and clusters far away from the jet axes are sensitive to CR effects, they are removed 
from the jets in alternative analyses by applying cones around the jet axes or by requiring a certain minmal
cluster energy.
The drawback of removing these clusters is an increasing statistical uncertainty.
Thus, studies have been performed to find the optimal cone opening angle or cluster energy cut.
As an example, in Figure~\ref{fig:crshift}~b) the decrease of the mass shift, $\Delta m_{\rm W}$, introduced by the CR model
is shown together with the increase of the statistical uncertainty, $\Delta\sigma_{\rm stat}$.
Also shown is the quadratic sum of both quantities, suggesting an optimal energy cut value of about 2 GeV.
Applying this cut decreases the CR uncertainty to about 50 MeV.
Even after these improvements are applied, the uncertainty of the W mass measured in the four-jet channel 
will still be dominated by final state interaction effects.

\section{Results}
\begin{figure}
  \begin{center}
    \includegraphics[width=.45\textwidth]{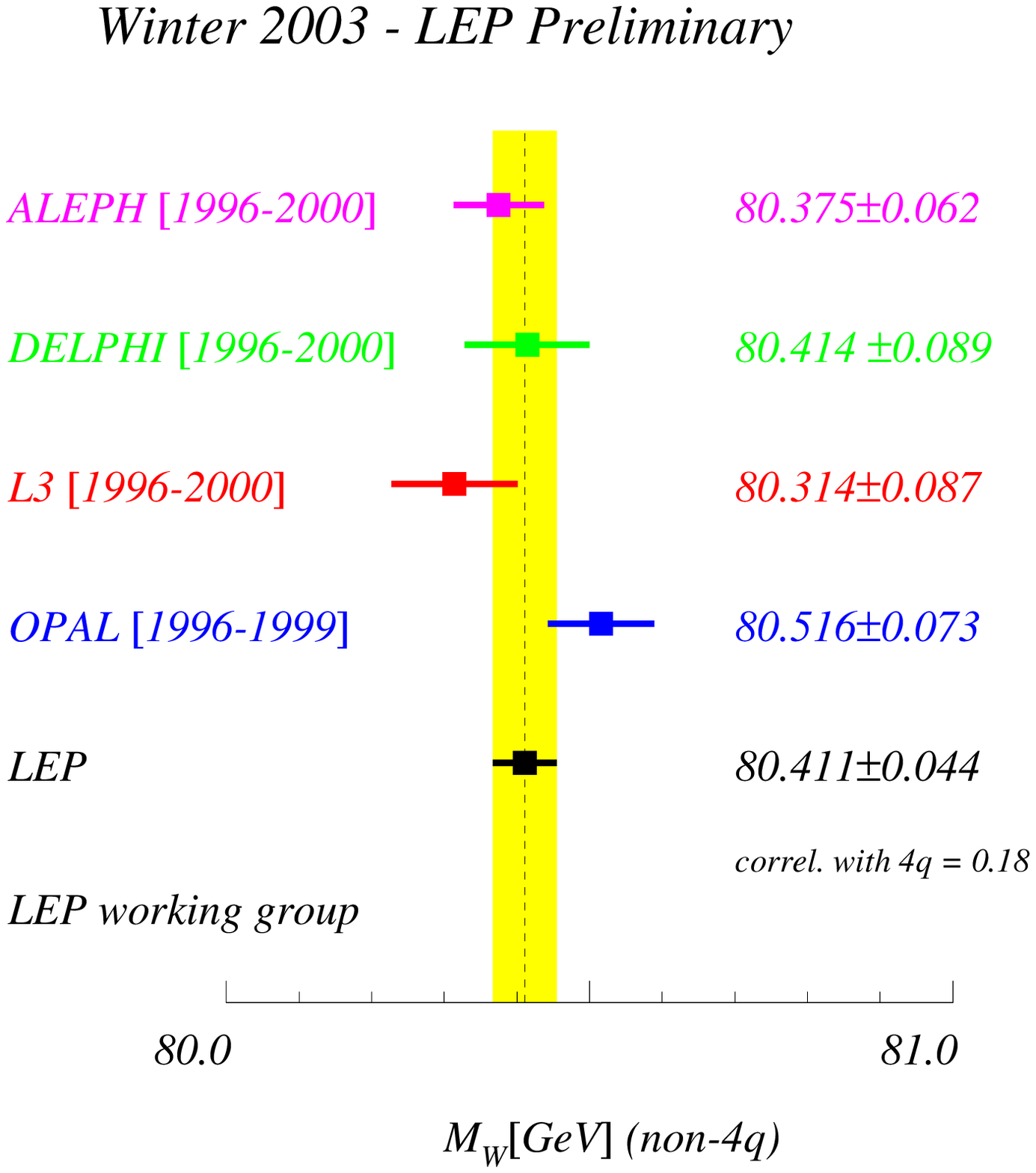}
    \hfill
    \includegraphics[width=.45\textwidth]{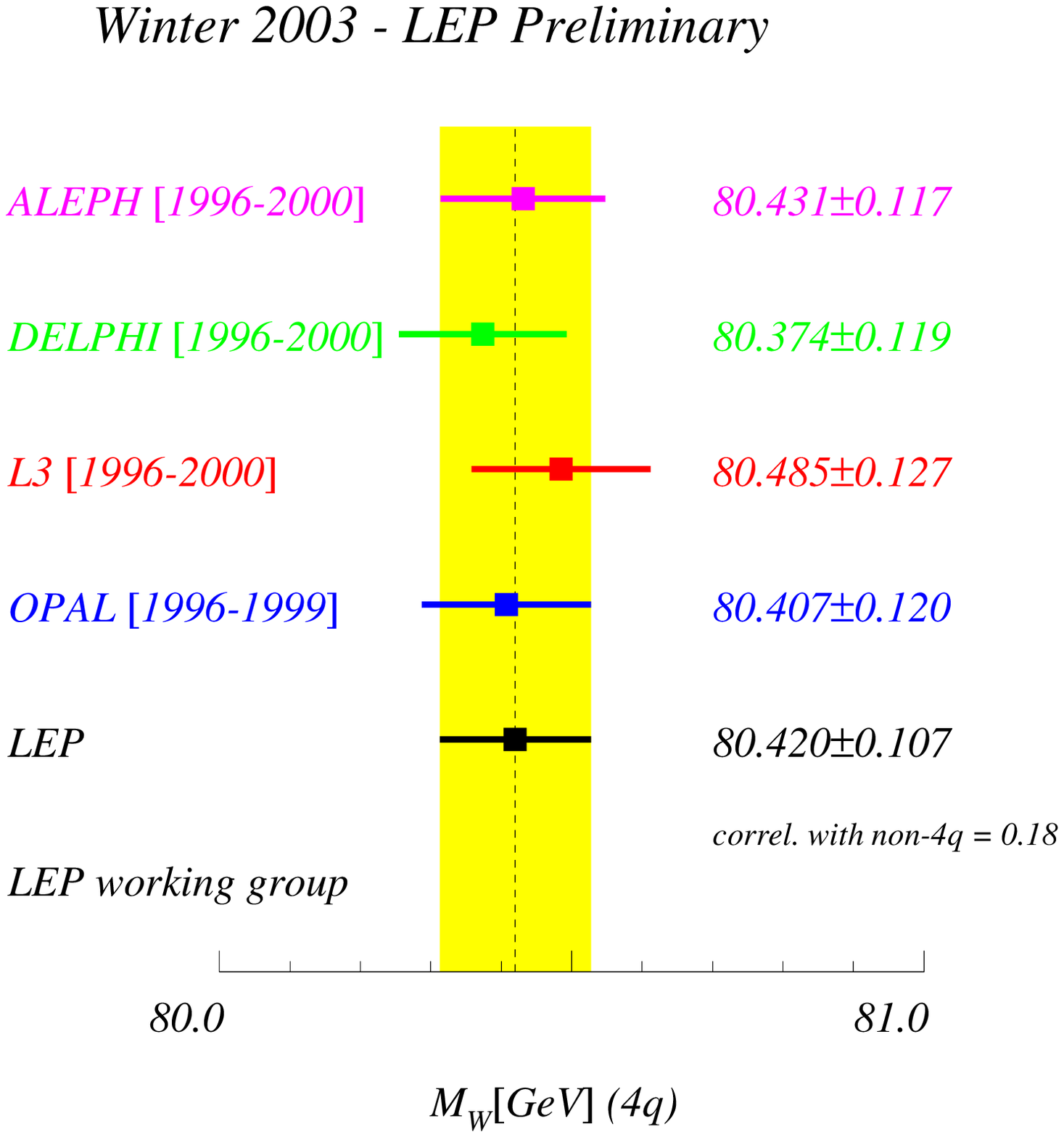}
  \end{center}
  \vspace{-22pt}
  a) \hspace{.49\textwidth}b)
  \vspace{22pt}
  \caption{
    Preliminary results of the measurement of the W boson mass.
    a) Measurements in the $\ell\nu\ell\nu$ an $\rm qq\ell\nu$ channels,
    b) measurements in the four-jet channel.
  }
  \label{fig:mw1}
\end{figure}

\begin{figure}
  \begin{center}
    \includegraphics[width=0.45\textwidth]{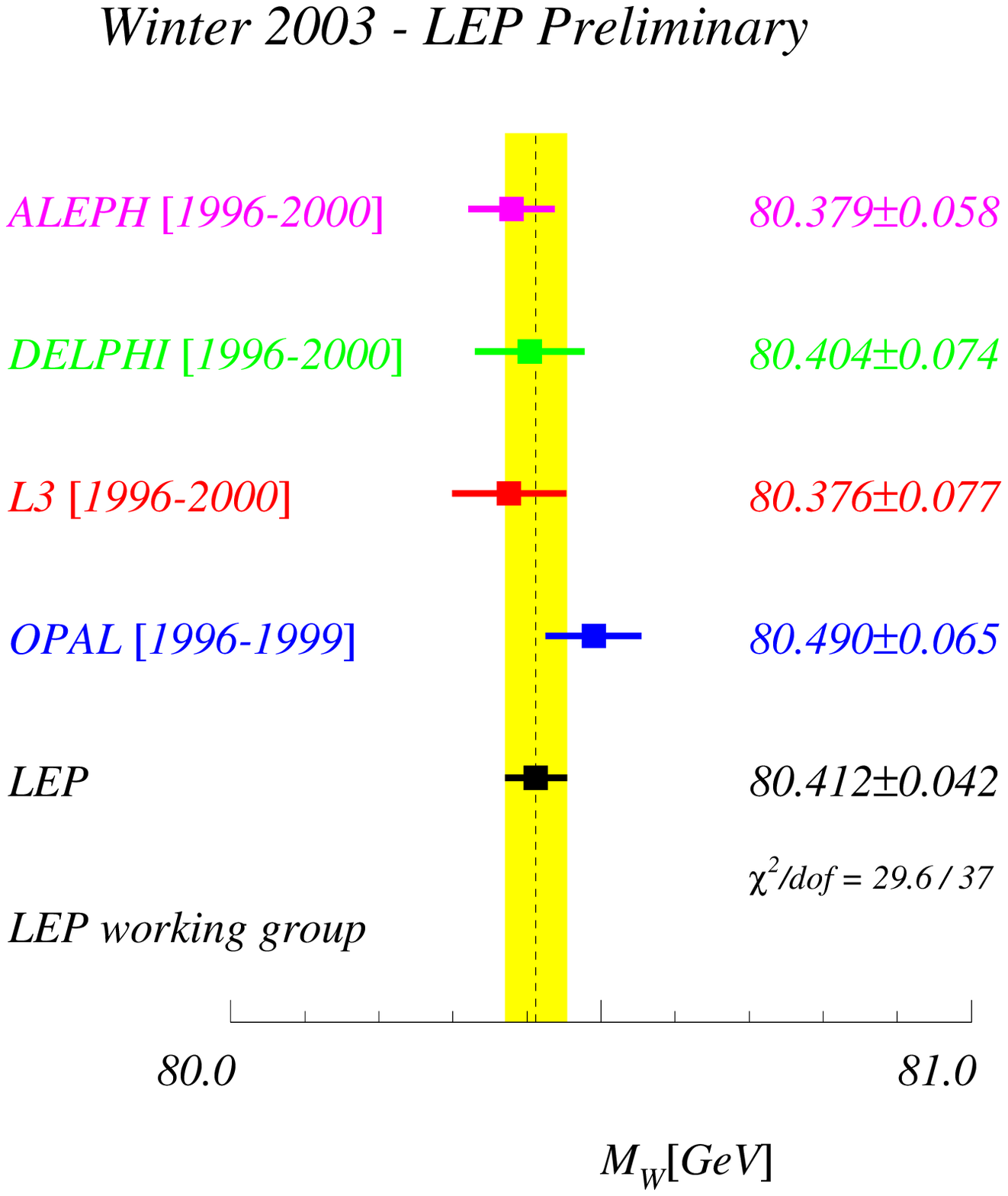}
    \hfill
    \includegraphics[width=0.45\textwidth]{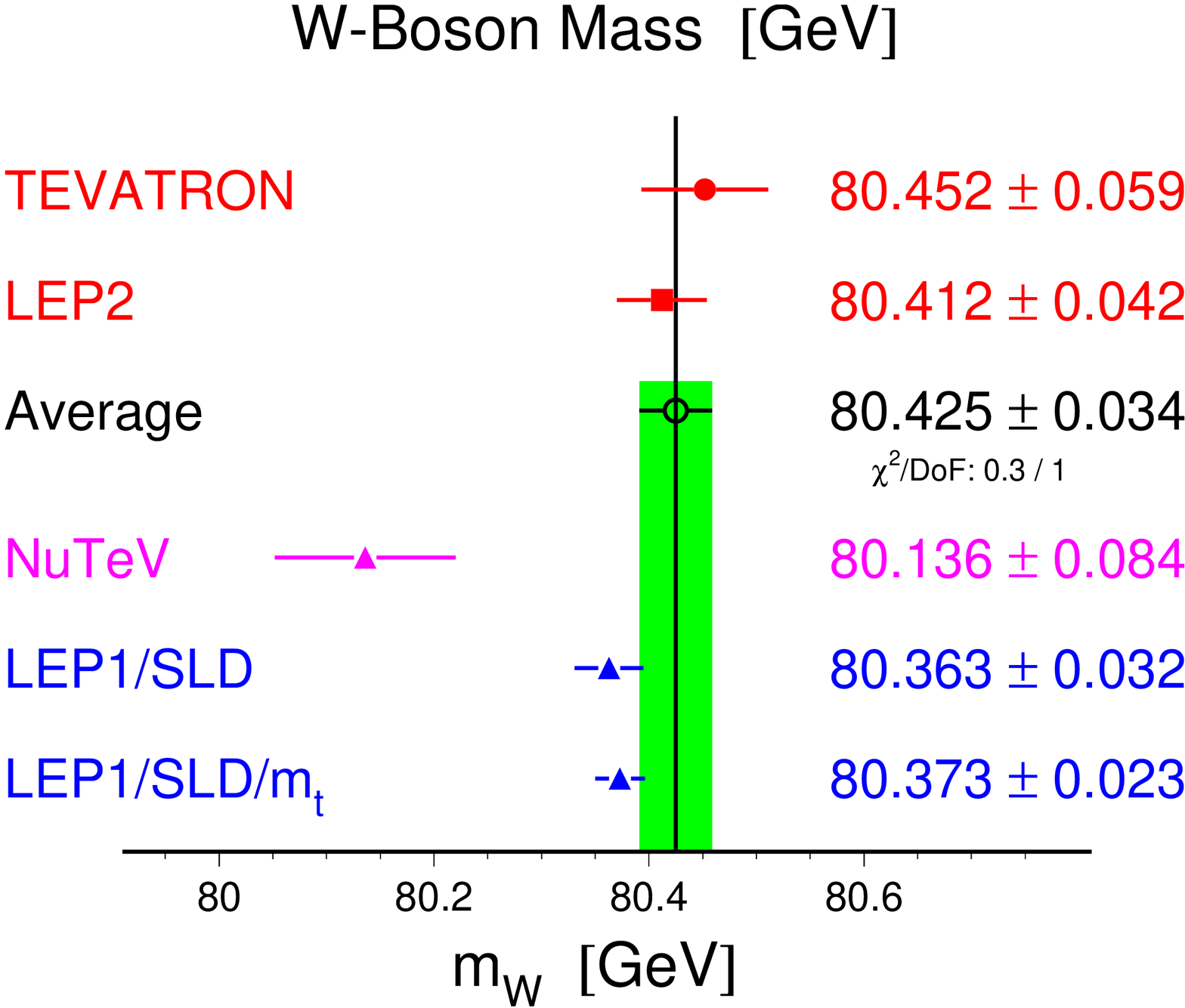}
  \end{center}
  \vspace{-22pt}
  a) \hspace{.49\textwidth}b)
  \vspace{22pt}
  \caption{
    a) Combination of measurements of the W boson mass in all channels for each LEP experiment and 
    combination of the four experiments.
    b) Comparison with other results.
  }
  \label{fig:mwall}
\end{figure}

The latest preliminary combination of W-boson mass measurements by the four LEP experiments~\cite{lepwcombi} results in
\begin{equation}
  m_{\rm W}^{\rm non-4q} = 80.411 \pm 0.032~({\rm stat.})\pm 0.030~({\rm syst.})~{\rm GeV}
\end{equation}
for the $\ell\nu\ell\nu$ and $\rm qq\ell\nu$ channels and
\begin{equation}
  m_{\rm W}^{\rm 4q} = 80.420 \pm 0.035~({\rm stat.})\pm 0.101~({\rm syst.})~{\rm GeV}
\end{equation}
for the qqqq channel.
The results of the individual experiments are shown in Figure~\ref{fig:mw1}.
Combining the two channels yields a value of
\begin{equation}
  m_{\rm W}^{\rm ffff} = 80.412 \pm 0.029~({\rm stat.})\pm 0.031~({\rm syst.})~{\rm GeV}.
\end{equation}
The combined result and a comparison with results from other experiments is shown in Figure~\ref{fig:mwall}.
Good agreement is observed between the LEP result and the direct measurement at $\rm p\bar{p}$ colliders and the
indirect measurements from LEP I and SLD.
The NuTeV results shows a deviation of about 2.9~$\sigma$ from the LEP result.
Recently it was suggested that the systematic uncertainties in the NuTeV analysis may be 
underestimated~\cite{Diener:2003ss}.

In a fit similar to the one used for the extraction of the W boson mass, the W boson width can be extracted.
The combination of all energies, channels, and experiments yields a value of
\begin{equation}
  \Gamma_{\rm W} = 2.150 \pm 0.091~{\rm GeV}.
\end{equation}
In this fit, $m_{\rm W}$ and $\Gamma_{\rm W}$ are treated as independent parameters.

\section{Electroweak fit}
\begin{figure}[t]
  \begin{center}
    \begin{minipage}{0.45\textwidth}
      \includegraphics[width=\textwidth]{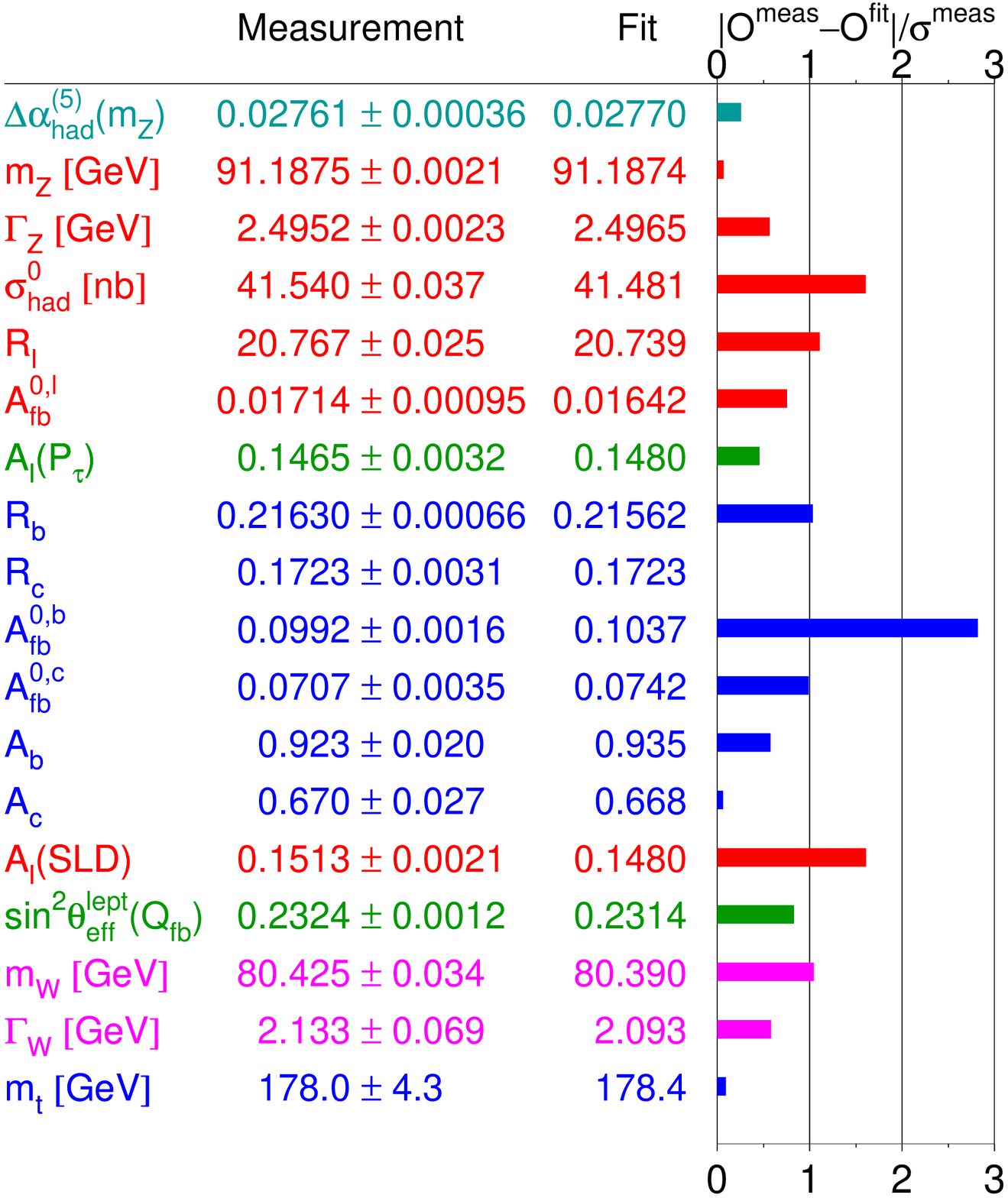}

      \vspace{-33pt}
    \end{minipage}
    \hfill
    \begin{minipage}{.45\textwidth}
      \includegraphics[width=\textwidth]{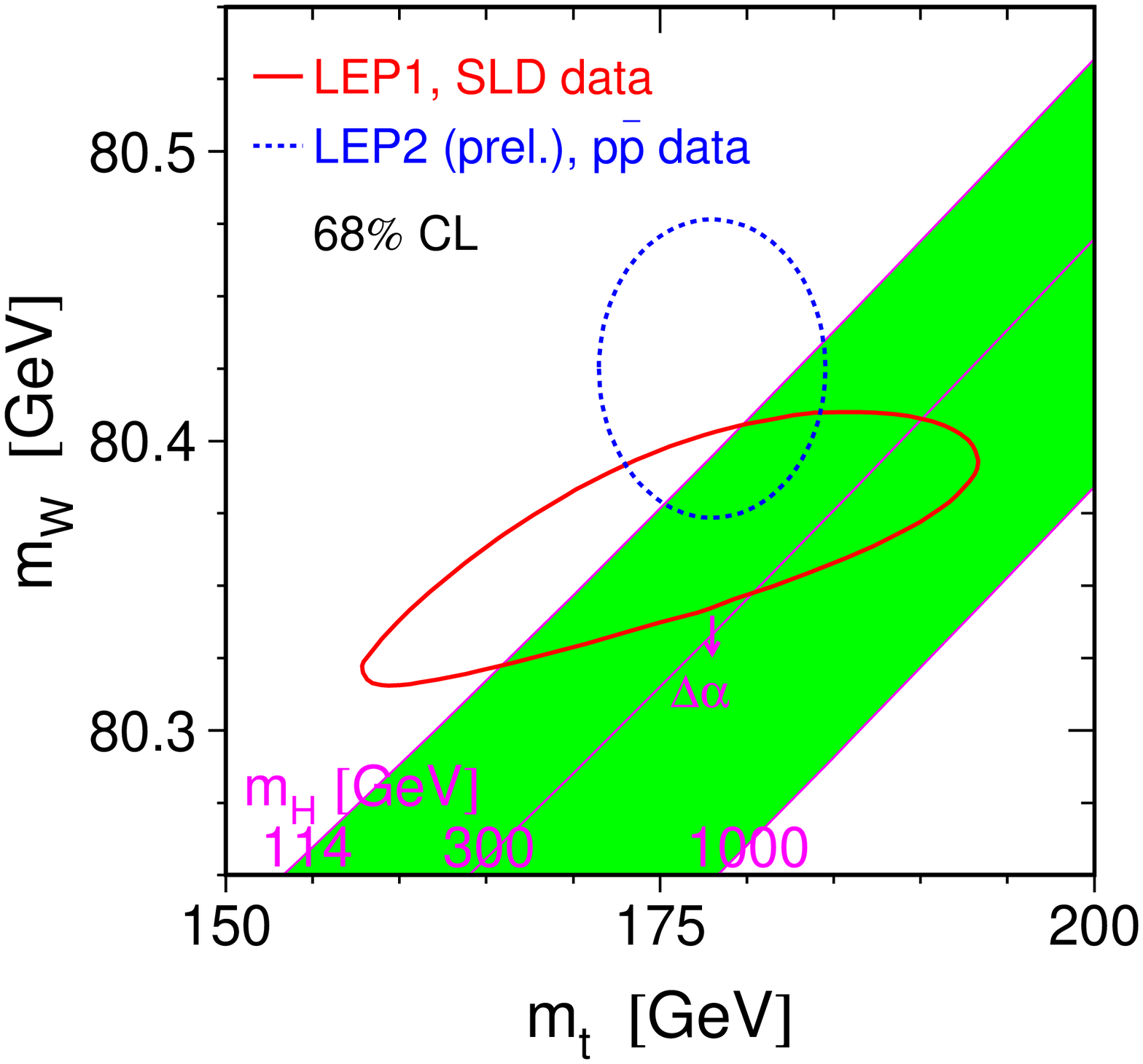}
    \end{minipage}
  \end{center}
  \vspace{-22pt}
  a) \hspace{.49\textwidth}b)
  \vspace{22pt}
  \caption{
    a) Input values for the electroweak fit, together with the individual fit results and the pull ofeach measure.
    b) 68\% CL contours for direct and indirect measurements of $m_{\rm W}$ and $m_{\rm t}$.
    The band displays predictions for $m_{\rm W}$ and $m_{\rm t}$ for different Higgs masses, using the precise measurement of $G_{F}$.
  }
  \label{fig:ewfit}
\end{figure}
The electroweak fit~\cite{lepcombi} combines measurements from a large number of experiments to test the consistency of the
Standard Model and to extract the parameter of the model that has not yet been measured directly, the mass of the
Higgs boson.
The input values for this fit are displayed in Figure~\ref{fig:ewfit}~a) together with the fit results and the pull of each 
measure.
With respect to Summer 2004, there have been three updates: 
Version 6.41 of the {\sc ZFitter} program~\cite{zfitter} is now used, 
the Z-pole heavy flavour combination has been updated~\cite{lepcombi},
and the value of $\Delta\alpha_{\rm had}^{(5)}(s)$ has slightly changed to $\Delta\alpha_{\rm had}^{(5)}(s) = 0.02758\pm 0.00035$~\cite{vacpol}.
In the fit and in Figure~\ref{fig:ewfit}~a), the old value is used, but the new result does not change the fit results.

The fit shows that there is agreement between the Standard Model predictions and the experimental measurements though
the $\chi^{2}$ of the fit has increased to 18.3 at 13 degrees of freedom using the latest experimental results.
The fit probability has therefore decreased to 14.7\% (compare to 26.1\% from Summer 2004).
The preferred mass for the Standard Model higgs boson has increased to
\begin{equation}
  m_{\rm H} = 126^{+73}_{-48}~{\rm GeV}.
\end{equation}

Figure~\ref{fig:ewfit}~b) displays the 68\% Confidence Level contours for measurements of $m_{\rm W}$ and the top quark mass, $m_{\rm t}$.
Shown are the predictions from indirect measurements at LEP I and SLD and for the direct measurements at LEP2 and $\rm p\bar{p}$ colliders.
Agreement between both measurements can be seen.
The band represents the Standard Model predictions for $m_{\rm W}$ and $m_{\rm t}$ for different masses of the 
Higgs boson using the precise value of the Fermi coupling constant, $G_{F}$.

\section{Conclusions and Outlook}
The analyses measuring the W boson mass are progressing well.
Final results are expected for Summer 2005, with an expected W mass uncertainty of about 35 MeV for the combination 
of the four LEP experiments.

The current preliminary value of the W-boson mass is
\begin{equation}
  m_{\rm W} = 80.412 \pm 0.029~({\rm stat.})\pm 0.031~({\rm syst.})~{\rm GeV}.
\end{equation}

The electroweak fit prefers a Higgs mass of
\begin{equation}
  m_{\rm H} = 126^{+73}_{-48}~{\rm GeV},
\end{equation}
with a fit probability of 14.7\%.

\end{document}